\documentclass{mn2e}

\usepackage{psfig,subfigure,mncite}

\newcommand{\bvec}[1]{\mbox{\boldmath ${#1}$}}

\begin{document}

\title[X-ray coronal structure]
{Inferring X-ray coronal structures from Zeeman-Doppler images}

\author
[M. Jardine, K. Wood, A. Collier Cameron, J.-F. Donati \& D.H. Mackay]
{M. Jardine$^1$, 
\thanks{E-mail: moira.jardine@st-and.ac.uk}
K. Wood$^1$, A. Collier Cameron$^1$, J.-F. Donati$^2$ \& D.H. Mackay$^3$ 
\\
$^1$School of Physics and Astronomy, Univ.\ of St~Andrews, 
St~Andrews, 
Scotland KY16 9SS \\
$^2$Laboratoire d¹Astrophysique, Observatoire Midi-Pyr\'en\'ees, 14 
Av. E. Belin, F-31400 Toulouse, France \\
$^3$School of Mathematics and Statistics, Univ.\ of St~Andrews, 
St~Andrews, Scotland KY16 9SS \\
\\
} 

\date{Received; accepted 2001}

\maketitle

\begin{abstract} 
We have modelled the X-ray emission from the young rapid rotator AB
Doradus (P$_{\rm rot}$ = 0.514 days) using as a basis Zeeman-Doppler
maps of the surface magnetic field.  This allows us to reconcile the
apparently conflicting observations of a high X-ray emission measure
and coronal density with a low rotational modulation in X-rays.  The
technique is to extrapolate the coronal field from the surface maps by
assuming the field to be potential.  We then determine the coronal
density for an isothermal corona by solving hydrostatic equilibrium
along each field line and scaling the plasma pressure at the loop
footpoints with the magnetic pressure.  We set the density to zero
along those field lines that are open and those where at any point
along their length the plasma pressure exceeds the magnetic pressure. 
We then calculate the optically thin X-ray emission
measure and rotational modulation for models with a range of coronal
densities.  Although the corona can be very extended, much of the
emission comes from high-latitude regions close to the stellar
surface.  Since these are always in view as the star rotates, there is
little rotational modulation.  We find that emission measures in the
observed range of $10^{52.8} - 10^{53.3}$cm$^{-3}$ can be reproduced
with densities in the range $10^{9}-10^{10.7}$cm$^{-3}$ for coronae at
temperatures of $10^{6}-10^{7}$K.

\end{abstract}

\begin{keywords}
 stars: activity -- 
 stars: imaging --
 stars: individual: AB Dor --
 stars: rotation --  
 stars: spots
\end{keywords}

\section{Introduction}

Studies of stellar coronae have been invigorated recently by the
wealth of results from Chandra and XMM-Newton.  Some previous
observations with ROSAT and ASCA had suggested that magnetically
active stars have coronae whose hotter gas is more extended than the
cooler gas (e.g. \pcite{gudel95}).  Eclipse mapping of binaries also
showed stars with extended coronal emission, of which the hotter
component suffered less rotational modulation
\cite{white90}.  This was not, however a universally
accepted view (see \pcite{singh96,siarkowski96,giampapa96}) and the physical
extent and location of the magnetic loops producing this emission was
very much in question \cite{jeffries98}.  One of the most challenging
results however came from line ratio studies that suggested that
coronal densities in the binaries Capella, $\sigma$ Gem and 44i Bootis
are very high, perhaps up to $10^{13}$cm$^{-3}$
\cite{dupree93,schrijver95,brickhouse98}.  More recent observations of
Capella by FUSE, Chandra and XMM-newton have also indicated high
densities \cite{audard01,mewe01,young01} (see also \pcite{gudel01} for
XMM-Newton results for a range of stars).

These high densities are problematical because, it is believed, such
plasmas can only be confined in small, compact, solar-like loops. 
These loops cannot explain the lack of rotational modulation, unless
they are located at very high latitudes and are always in view as the
star rotates or as it is eclipsed by a binary companion.  Doppler
images of active stars do indeed often show polar or high-latitude
spots (see \pcite{strassmeier96table} for a review), but they also
typically show spots at all other latitudes.  For AB Dor in
particular, Zeeman-Doppler images are also available
{\cite{donati95,donati97}.  These show flux at the kilogauss level at
all latitudes.  Measurements of coronal densities for AB Dor are also
high, ranging from the $3\times10^{10}$cm$^{-3}$ derived from
XMM-Newton observations \cite{gudel01} to the $3 \times
10^{9}$cm$^{-3}$ in quiescence and $10^{12}$cm$^{-3}$ during flaring
derived from modelling a flare decay observed with BeppoSAX
\cite{maggio2000}.  This last observation is particularly interesting
as it follows the flare decay over a timescale longer than a
rotation cycle.  There was no rotational eclipse or modulation of the
emission, and modelling of the flare decay phase suggested that the
loop structure was small, with a maximum height of only about
$0.3$R$_{\star}$.  This suggests that the flaring loop or loops must
have originated at latitudes above 60$^{\circ}$ (the stellar
inclination) in order to remain in view.  Observations of directed
radio emission from AB Dor \cite{lim94} also place the emitting
regions at high latitudes.

Simply placing the X-ray emitting loops close to the pole may not be
the whole solution, however.  Emission measures for these stars are
typically very high.  Values determined for AB Dor at temperatures
from $10^{6}$K to above $10^{7}$K based on observations with EXOSAT
\cite{cameron88xray}, ASCA/EUVE \cite{mewe96}, XMM-Newton
\cite{gudel01} and HST \cite{vilhu2001} have upper and lower limits of
$10^{52.8} - 10^{53.3}$cm$^{-3}$ when summed over all temperature
bins.  These emission measures require a large volume even at the
observed high densities. Any model must also explain any observed
rotational modulation, or the lack of it.  \scite{kurster97} presented
a long-term study of the X-ray emission from AB Dor and concluded that
a small rotational modulation at the level of $5\%-13\%$ was present. 
Coronal models must also be able to explain the presence of very
massive prominences often observed in active stars.  In the case of AB
Dor, these are typically observed to be trapped in the corona at some
3-5 R$_{\star}$ from the rotation axis
\cite{cameron89cloud,cameron89eject}.  In order for these prominences
to form, there must be sufficient material at these heights to sustain
the radiative instability thought to trigger their formation.

\begin{figure}

\psfig{figure=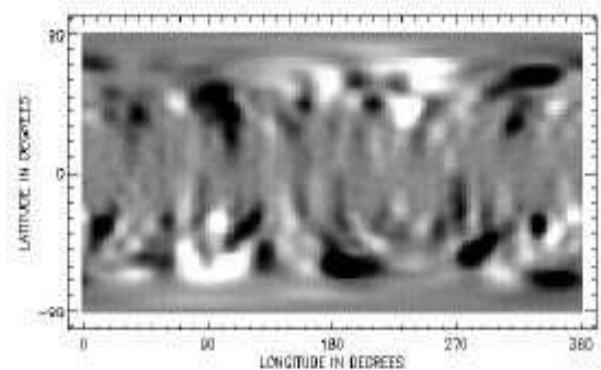,width=8cm,height=5cm}

 \caption{A map of the surface radial magnetic field of AB Dor.  White
 represents -800G and black represents 800G. Since AB Dor is inclined
 at 60$^{\circ}$ to the observer, Zeeman-Doppler images provide only
 limited information in the lower hemisphere.  In order to compensate
 for this, we have generated this combined surface map, with the 1995
 map in the upper hemisphere and the 1996 map in the lower hemisphere. 
 The observable hemisphere of AB Dor typically shows one longitude
 range of predominantly positive polarity, while the remaining range of
 longitudes is of predominantly negative polarity.  We have chosen the
 alignment of the 1995 and 1996 maps such that they are antisymmetric
 about the equator, with the predominantly positive polarity regions
 180 degrees apart in longitude in the two hemispheres.  }

  \label{mixed180_icont0}
 
\end{figure}

The aim of this paper is to use the Zeeman-Doppler maps of the 
surface magnetic field of AB Dor to extrapolate the coronal field and 
hence, by assuming hydrostatic equilibrium, to determine the coronal 
density and X-ray emission. By varying the base density and 
calculating the emission measure and the rotational modulation, we 
can determine if it is possible to reconcile the observations of high 
densities and high emission measures with the low rotational modulation.

\section{Field extrapolation}

The method of extrapolating the coronal field has been described in
\scite{jardine01structure} and will not be repeated in detail here. 
We use the source surface method pioneered by \scite{altschuler69} and
a code originally developed by \scite{vanballegooijen98}.  Briefly, we
write the magnetic field $\bvec{B}$ in terms of a flux function $\Psi$
such that $\bvec{B} = -\bvec{\nabla} \Psi$ and the condition that the
field is potential ($\bvec{\nabla}\times\bvec{B} =0$) is satisfied
automatically.  The condition that the field is divergence-free then
reduces to Laplace's equation $\bvec{\nabla}^2 \Psi=0$ with solution 
in spherical co-ordinates $(r,\theta,\phi)$
\begin{equation}
 \Psi = \sum_{l=1}^{N}\sum_{m=-l}^{l} [a_{lm}r^l + b_{lm}r^{-(l+1)}]
         P_{lm}(\theta) e^{i m \phi},
\end{equation}
where the associated Legendre functions are denoted by $P_{lm}$.  The
coefficients $a_{lm}$ and $b_{lm}$ are determined by imposing the
radial field at the surface from the Zeeman-Doppler maps and by
assuming that at some height $R_s$ above the surface the field becomes
radial and hence $B_\theta (R_s) = 0$.  This second condition models
the effect of the plasma pressure in the corona pulling open field
lines to form a stellar wind.  Since large slingshot prominences are
observed on AB Dor mainly around the co-rotation radius which lies at
$2.7R_\star$ from the rotation axis, we know that a significant
fraction of the corona is closed out to those heights and so we set
the value of $R_s$ to $3.4R_\star$.

We find that much of the corona is filled with open field
lines, and will therefore be dark in X-rays.  These open field lines
originate in two opposite-polarity, mid-latitude regions located
180$^{\circ}$ apart in longitude.  The closed field volume between
these two ``coronal holes'' forms a torus that extends almost over
both poles and whose axis connects the two ``open field'' longitudes. 

\section{Calculating the density}

In order to relate our model to the observations, we determine the
X-ray emission from the closed field regions.  As a first step, we
calculate the pressure structure of the corona assuming it to be
isothermal and in hydrostatic equilibrium.  Hence for a stellar
rotation rate $\omega$, the pressure at any point is $p=p_{0}e^{\int
g_{s}ds}$ where $g_{s} =( {\bf g.B})/|{\bf B}|$ is the component of
gravity (allowing for rotation) along the field and
\begin{equation}
g(r,\theta) = \left( -GM_{\star}/r^{2} + 
                     \omega^{2}r\sin^{2} \theta,
		     \omega^{2}r\sin \theta \cos\theta 
             \right).    
\end{equation}
 We note that for AB Dor, $\omega = 1.4\times10^{-4}$s$^{-1}$, a
 factor of 50 times greater than the corresponding solar value.  At
 the loop footpoints we scale the plasma pressure $p_{0}$ to the
 magnetic pressure such that $p_{0}(\theta,\phi)=R
 B^{2}_{0}(\theta,\phi)$ where $R$ is a constant.  The plasma pressure
 within any volume element of the corona is set to zero if the field
 line through that volume element is open.  In order to mimic the
 effect of a high gas pressure forcing closed field lines to open up,
 we also investigate the effect of setting the plasma pressure to
 zero if the plasma pressure is greater than the magnetic pressure
 i.e. where $\beta>1$.  From the pressure, we calculate the density
 assuming an ideal gas and determine the morphology of the optically
 thin X-ray emission by integrating along lines of sight through the
 corona.

\begin{figure}

	\psfig{figure=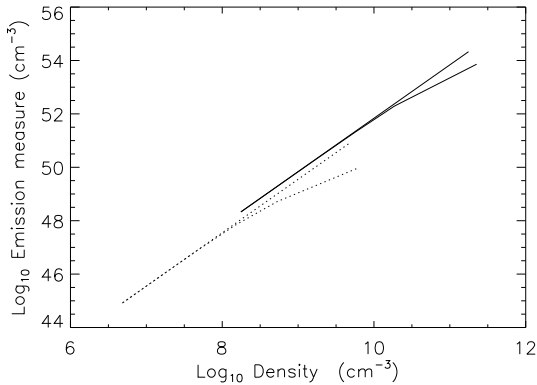,width=8.0cm} 

	\caption[]{Emission measure $\int n_{e}^{2}dV$ for an
	isothermal corona at a temperature of $10^{6}$K for solar
	maximum (solid line) and solar minimum (dotted line) as a
	function of the emission-measure weighted density.  In each
	case the branching of the curves at high densities shows the
	amount by which the emission is reduced if field lines where
	the plasma pressure exceeds the magnetic pressure are assumed
	to have been forced open.  }
	
    \label{em_sun}
\end{figure}

 
 This technique has been used for some time now to extrapolate the
 coronal field of the Sun.  Comparisons of the resulting morphology of
 the corona with LASCO C1 images \cite{wang97} show good agreement for
 much of the corona, except near the polar hole boundaries where the
 potential field approximation is likely to break down at the
 interface between open and closed field regions.  \scite{wang97}
 found the best agreement with observations was obtained with a
 scaling of $p_{0} \propto B^{0.9}_{0}$.  The question of the optimal
 scaling is one that is often addressed in the context of the heating
 of the solar corona (see \pcite{aschwanden2001} and references
 therein) and indeed recent work suggests that scaling laws developed
 on the basis of detailed solar observations may be extrapolated to
 more rapidly rotating stars \cite{schrijver2002}.  Given that our
 intent is not to discriminate between heating models, but rather to
 demonstrate that the observed surface field distributions are
 consistent with the X-ray observations, we opt for the simplest
 prescription that fits the data.  Our choice of $p_{0} \propto
 B^{2}_{0}$ ensures that regions of high field strength will emit
 strongly.  This naturally concentrates much of the emission at high
 latitudes.
 

\begin{figure}

	\psfig{figure=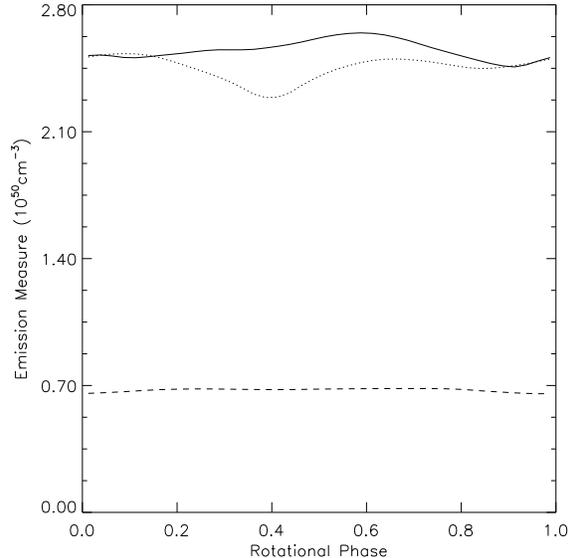,width=8.0cm} 

	\caption[]{Rotational modulation of the emission measure for
	three different surface maps.  The dashed curve is for the
	1995 dataset.  The other two curves are for a mixed dataset
	with the 1995 map in the upper hemisphere and the 1996 dataset
	in the lower hemisphere.  In each case the relative
	orientation of the two maps is different.  The dotted line is
	the case where the regions of predominantly the same polarity in
	each hemisphere are 180$^{\circ}$ apart in longitude; the
	solid line is the case shown in Fig.  \ref{mixed180_icont0}
	where they are at the same longitude.  }
	
    \label{rotmod}
\end{figure}


\begin{figure}
	        \label{em_dip}
		\psfig{figure=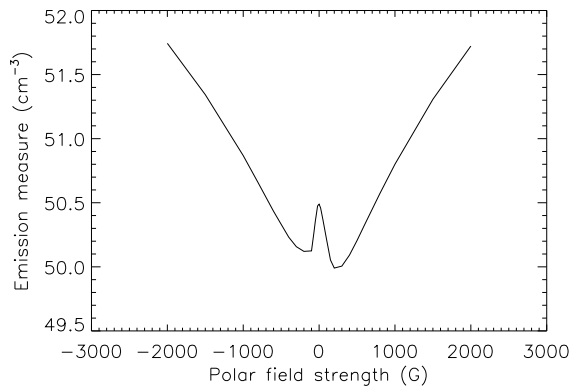,width=8.0cm}

		\caption[]{Emission measure for the combined surface
		map shown in Fig.  \ref{mixed180_icont0} with a dipole
		component of different strengths added.  }
		
\end{figure}


\begin{figure*}
	\def\subfigtopskip{4pt}
	\def\subfigbottomskip{4pt}
	\def\subfigcapskip{2pt}
	\centering
	\begin{tabular}{ccc}
    	        \subfigure[]{
			\label{128_linesonly} 			
			\psfig{figure=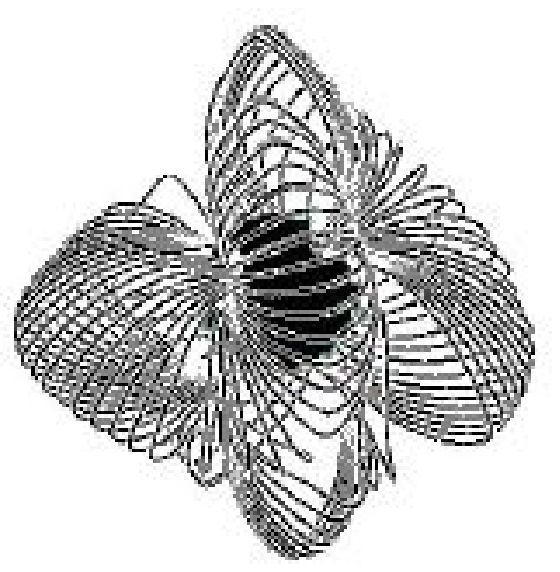,width=5.5cm}
			} &
		\subfigure[]{
			\label{nocutoff_image_1} 						
			\psfig{figure=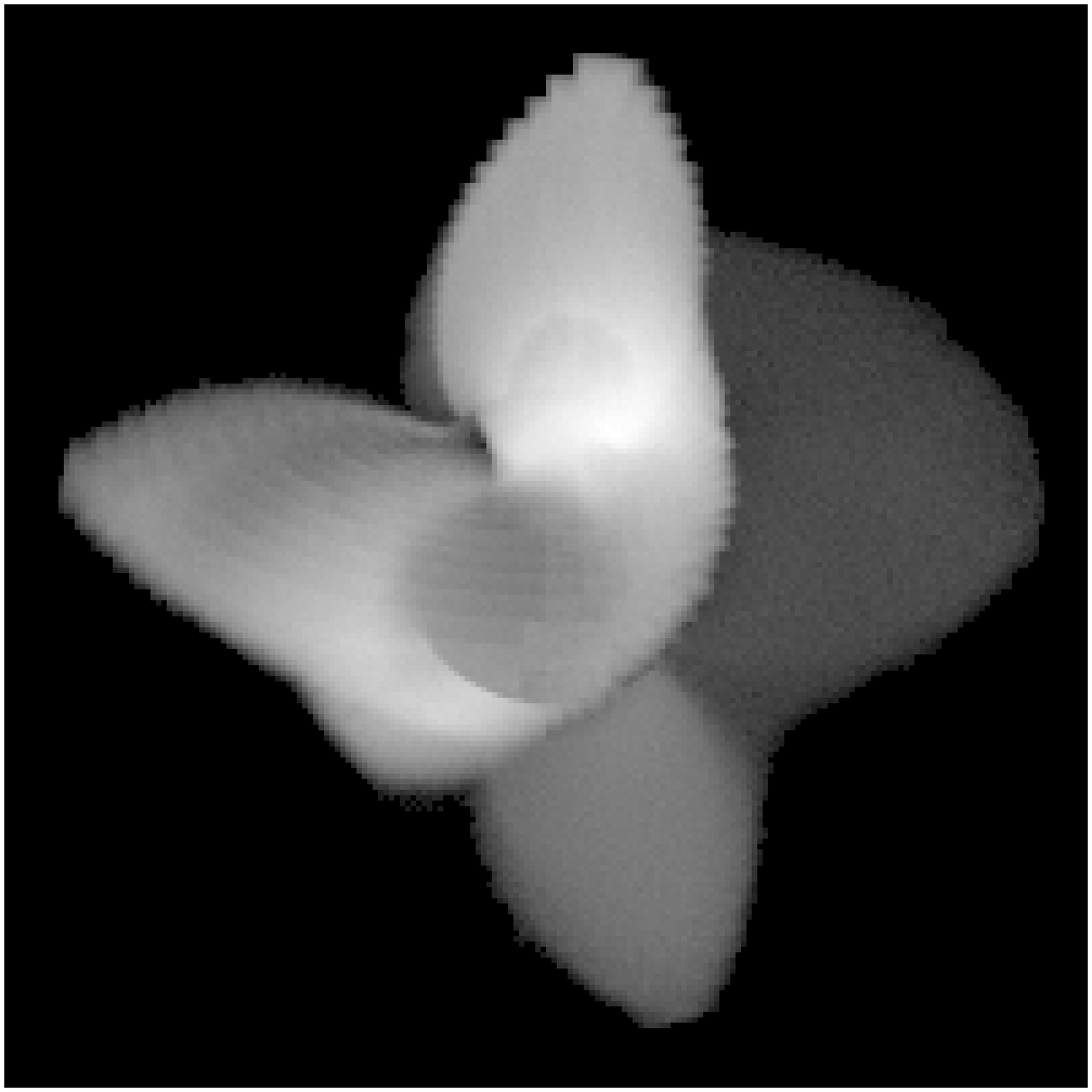,width=5.5cm}
			} &
		\subfigure[]{
			\label{images_lowT} 						
			\psfig{figure=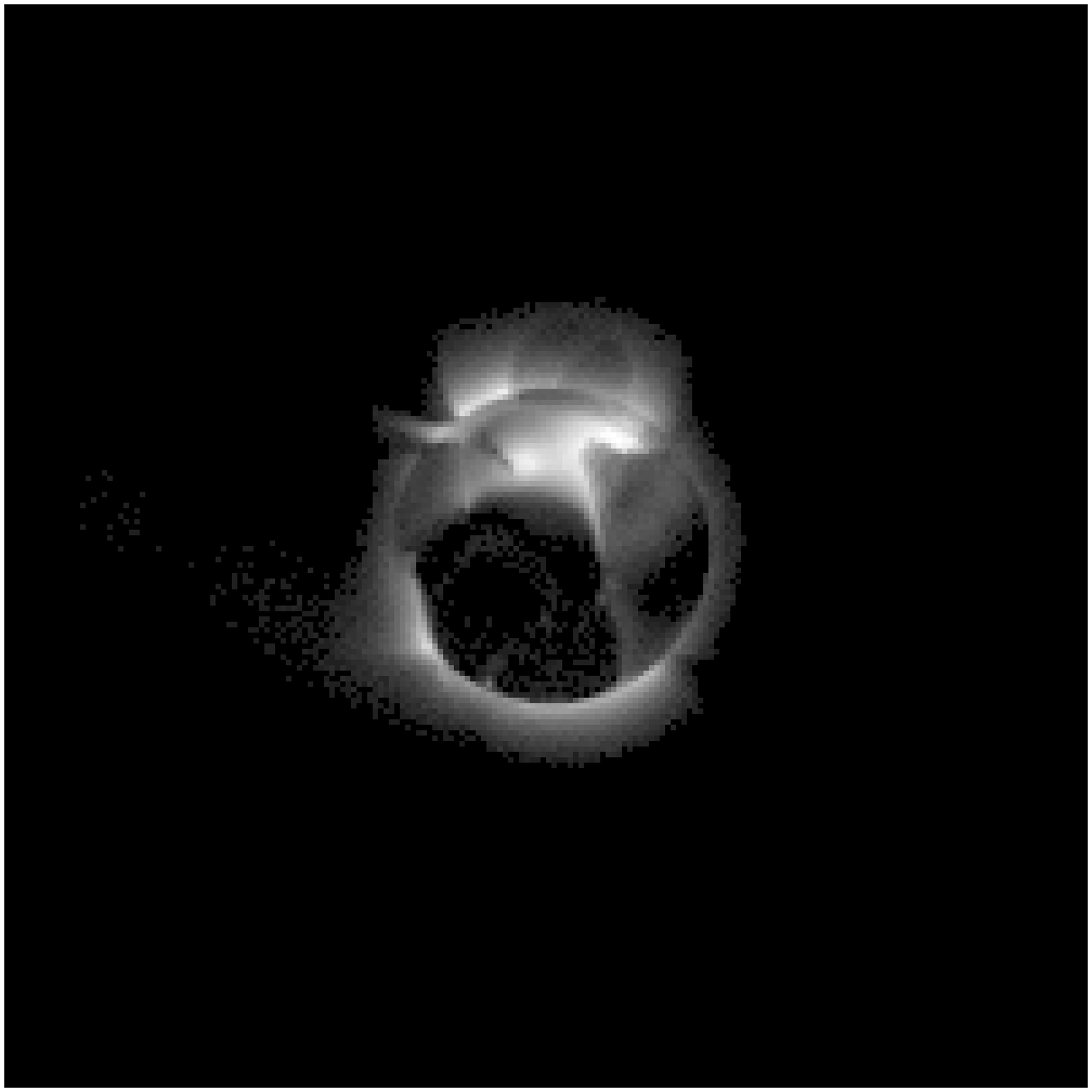,width=5.5cm}
			} \\
	\end{tabular} 
\caption[]{The effect of the coronal temperature on the emission
measure.  Fig.  \ref{128_linesonly} shows the magnetic structure of
the large-scale field based on the surface map shown in Fig. 
\ref{mixed180_icont0}.  The field is viewed from longitude
$128^{\circ}$.  Figs.  \ref{nocutoff_image_1} and \ref{images_lowT}
show the corresponding emission measure images on a squre-root stretch
at temperatures of 10$^{7}$K, and 10$^{6}$K respectively.  The
corresponding emission-measure weighted densities are $4\times
10^{8}$cm$^{-3}$ and $2\times 10^{8}$cm$^{-3}$.}
    \label{no_cutoff_images}
\end{figure*}	

 In order to test the model, we extrapolate the field from two solar
 magnetograms taken from the NSO/KP data archive, one from solar
 maximum (1992 Feb 1) and one from solar minimum (1996 April 8).  With
 a source surface set to 2.5 R$_{\odot}$ and a uniform temperature of
 $10^{6}$K we determine the total emission measure ($\int n_{e}^{2}
 dV$) for a range of values of the scaling constant $R$.  Although the
 density at the base of the corona is the parameter that we can vary
 with this model, this is not necessarily the density that is
 observed.  We therefore determine the emission-measure weighted
 density ($\int n_{e}^{3}dV / \int n_{e}^{2}dV$) and use this as a
 measure of the coronal density.  Fig.  \ref{em_sun} shows the
 resulting emission measure for the two magnetograms.  In each case,
 emission measures were calculated both with and without a cutoff for
 high pressures.  This causes the branching of the curves at high
 densities, with the lower branch being the case where a cutoff was
 imposed.  We do not expect this simple one-temperature model to
 reproduce the detailed changes in the magnitude and temperature
 distribution of the Sun's X-ray emission through its cycle. 
 Nonetheless, the overall morphology of the emission and the observed
 range of emission measures of $10^{49}-10^{50}$cm$^{-3}$ are easily
 reproduced and correspond to densities of the order of
 $10^{9}$cm$^{-3}$ \cite{peres2000}.
 
\section{Modelling the corona of AB Dor}

In the case of the Sun, resolved X-ray observations can be used to test
field extrapolation models.  For other stars however we generally have
only the emission measure and the rotational modulation to give us
insight into the field structure.  In this section we determine the
effect of the different assumptions of our model on these two
quantities.

\subsection{The effect of hidden flux}
Flux can be undetected by the Zeeman-Doppler method if it is located
in the unobservable hemisphere of the star, or in regions of the
stellar surface that are dark, such as the polar cap. 
\scite{jardine01structure} explored the effect of this hidden flux by
adding to the maps of the observable hemisphere either a polar field
(in the form of a global dipole) or the surface map for another year
added in to the hidden hemisphere (see Fig.  \ref{mixed180_icont0}). 
In all these cases the structure of the field (and in particular the
positions of the open field regions) is quite different.  These
differences are, however, unlikely to be apparent in the rotational
modulation and magnitude of the X-ray emission measure which is the
main observational signature of this field.

This is demonstrated in Fig.  \ref{rotmod} which shows this rotational
modulation for three different surface maps.  The lowest curve comes
from a surface map that has data only in the observable hemisphere. 
There is very little rotational modulation, since flux in the lower
hemisphere which might be eclipsed as the star rotates is missing. 
The upper two curves show the effect of having flux in both
hemispheres.  In one case, the predominantly positive polarity regions
in the two hemispheres are aligned, and in the other case they are
180$^{\circ}$ apart in longitude (this is the case shown in Fig. 
\ref{mixed180_icont0}).  There is little difference in the rotational
modulation of these two cases.

\begin{figure*}
	\def\subfigtopskip{4pt}
	\def\subfigbottomskip{4pt}
	\def\subfigcapskip{2pt}
	\centering
	\begin{tabular}{ccc}
		\subfigure[]{
			\label{lowp_image_1} 
			\psfig{figure=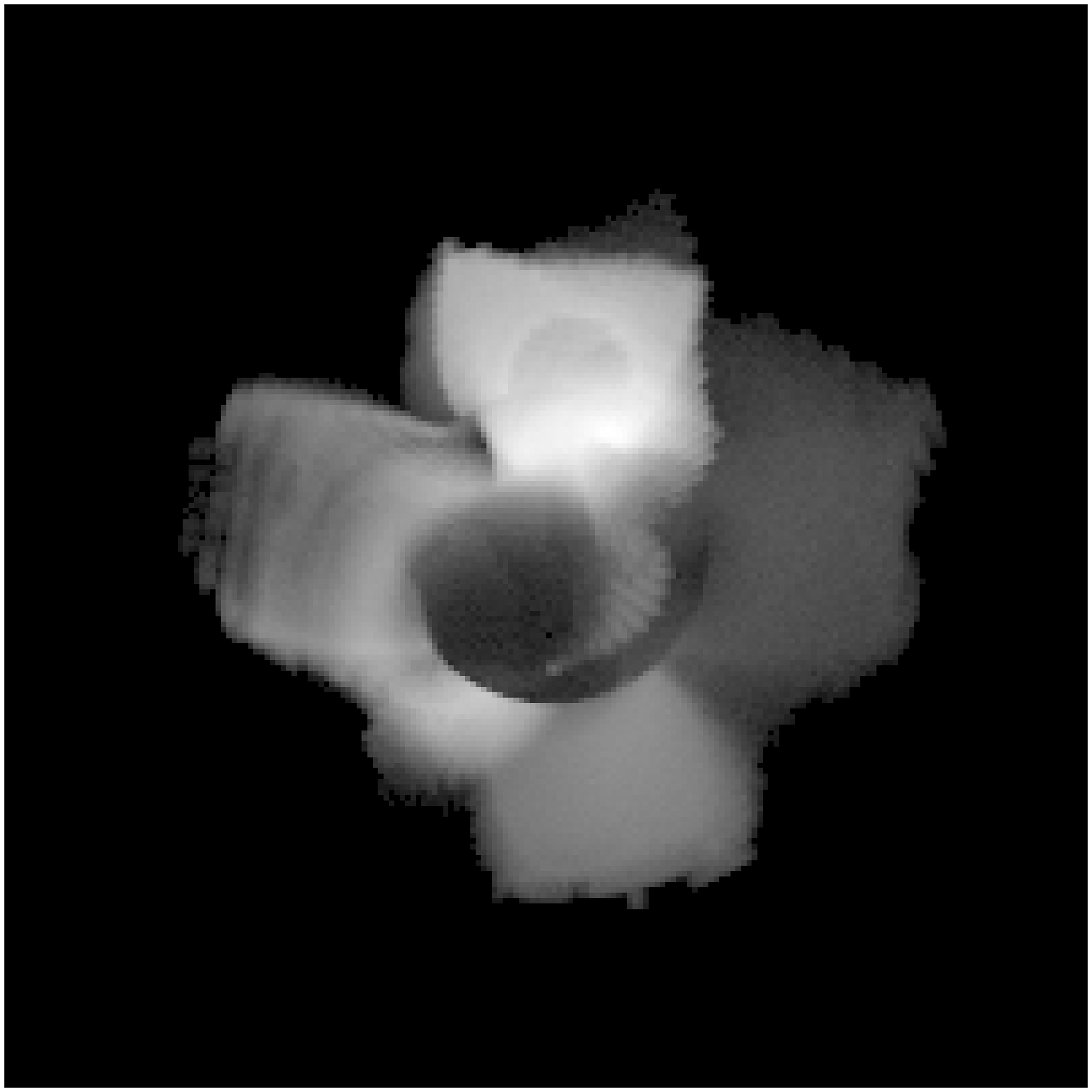,width=5.5cm}
			} &
		\subfigure[]{
			\label{lowp_image_2} 					
			\psfig{figure=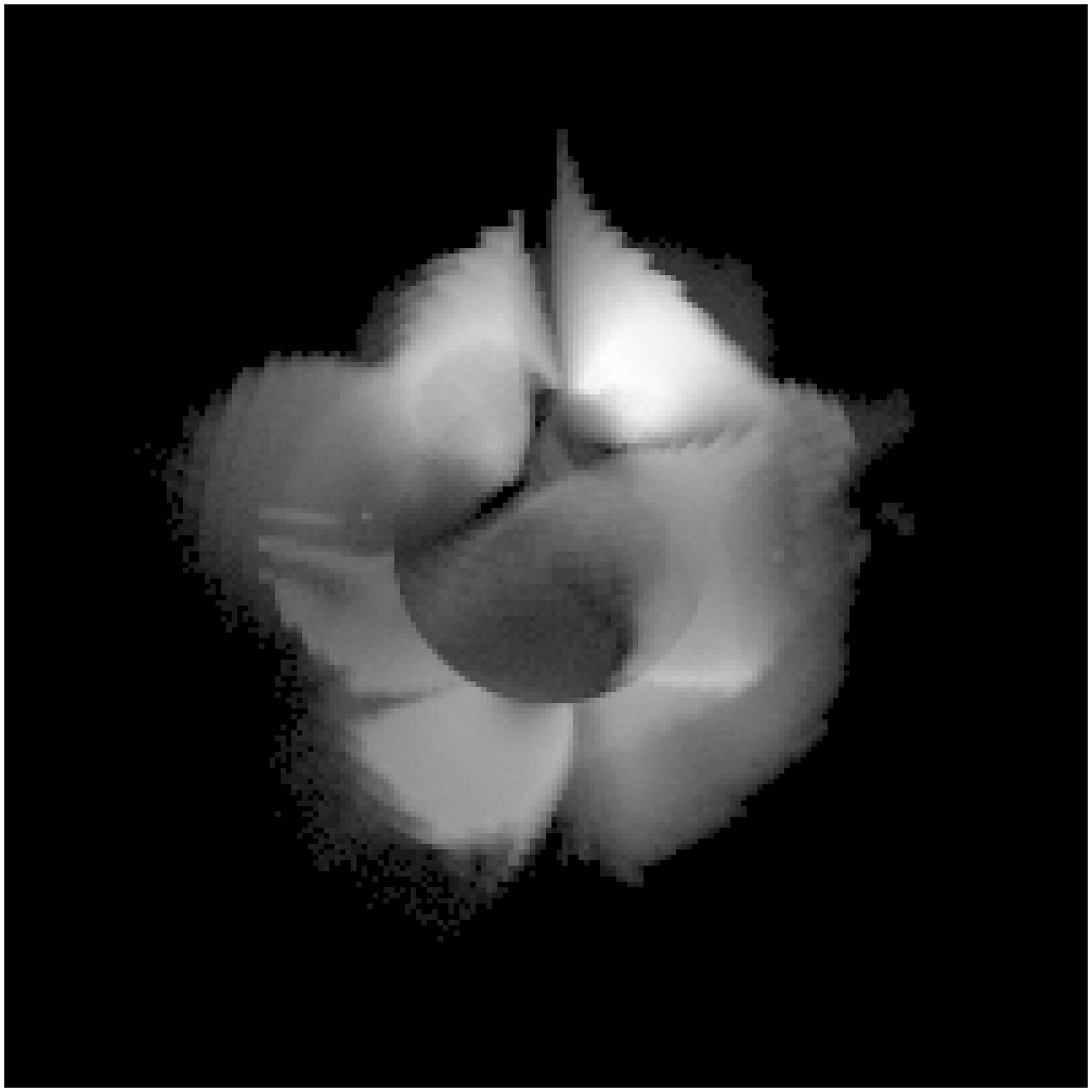,width=5.5cm}
			} &
		\subfigure[]{
			\label{lowpmod} 						
			\psfig{figure=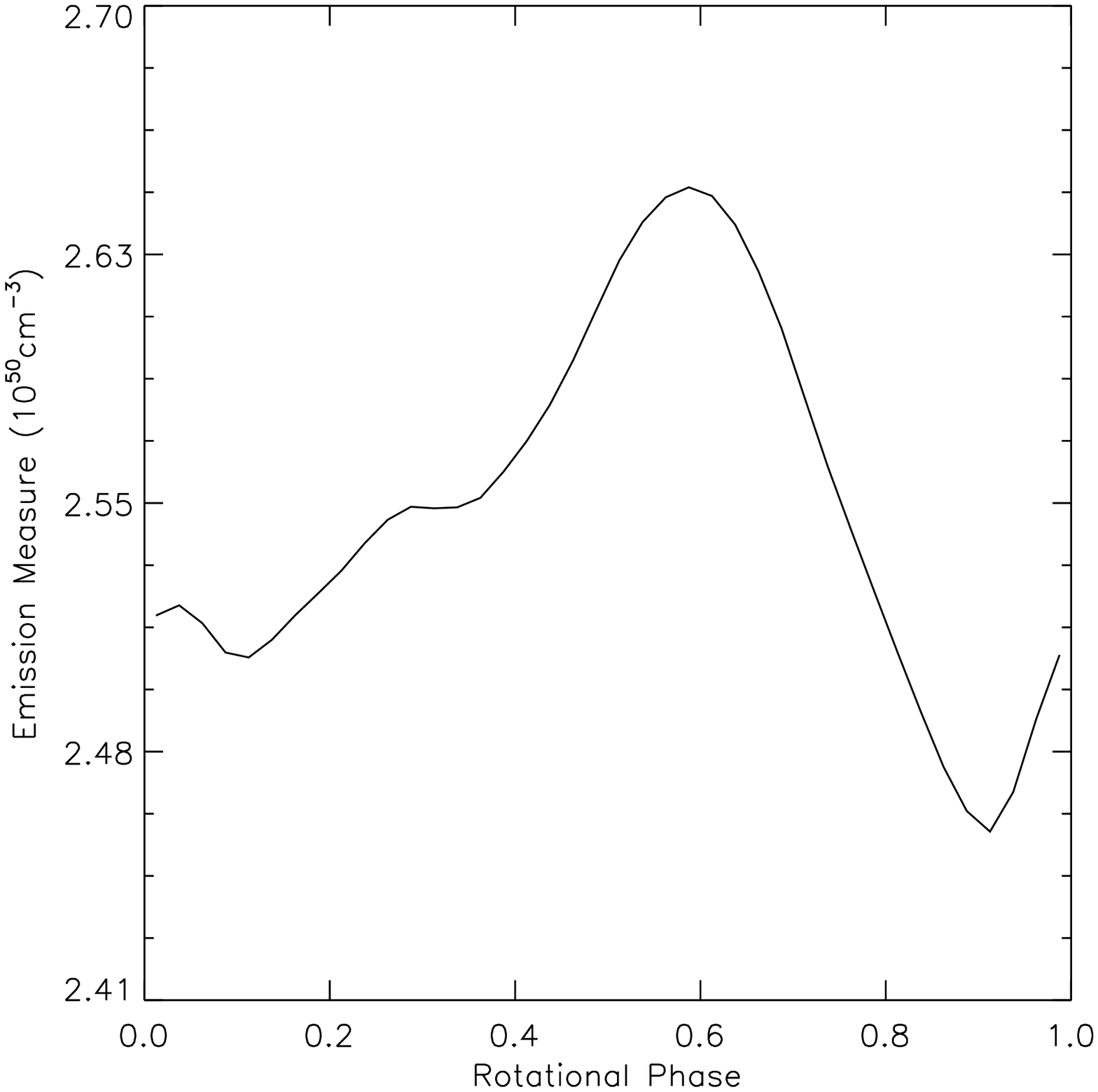,width=5.5cm}
			} \\
		\subfigure[]{
			\label{medp_image_1} 		
			\psfig{figure=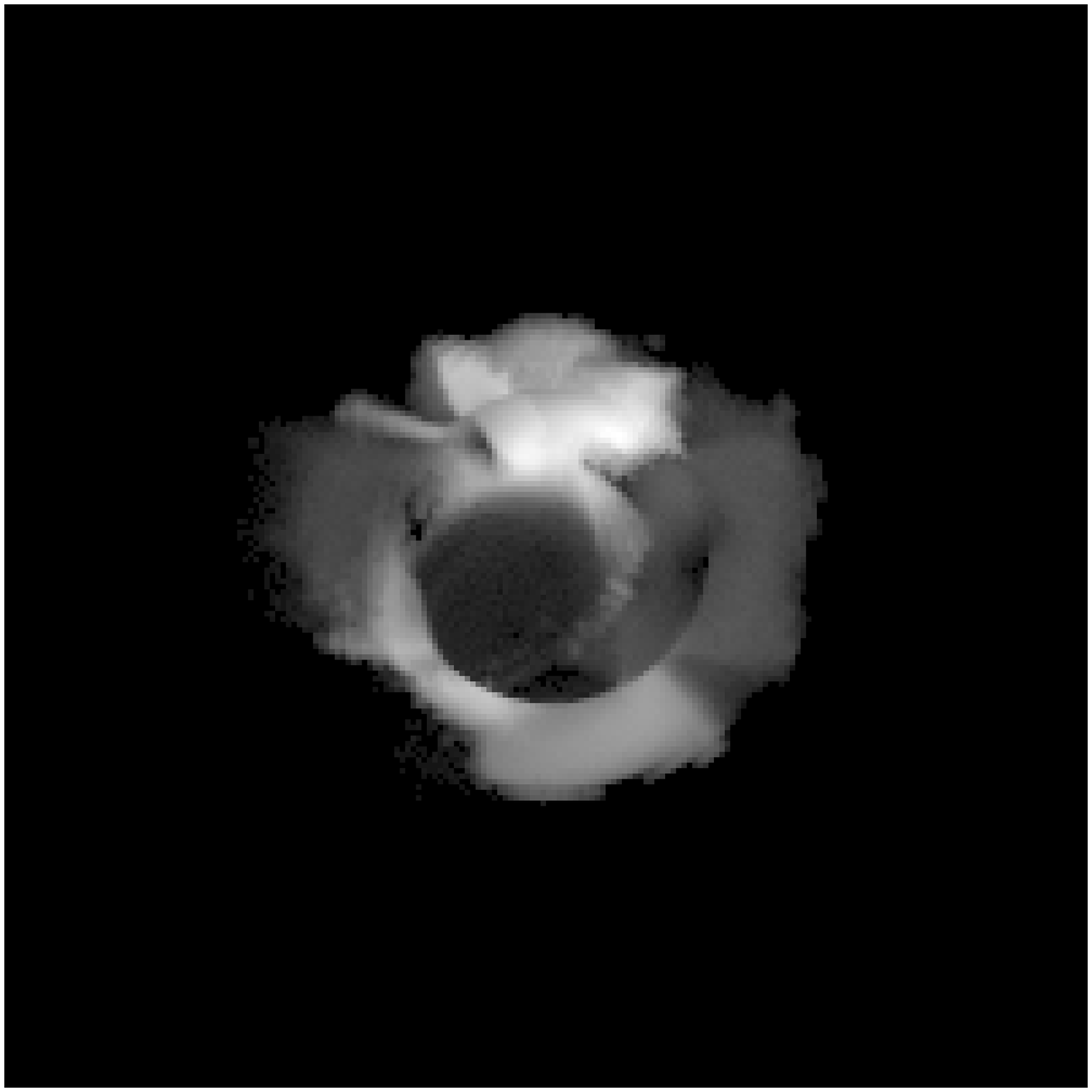,width=5.5cm}
			} &
		\subfigure[]{
			\label{medp_image_2} 
			\psfig{figure=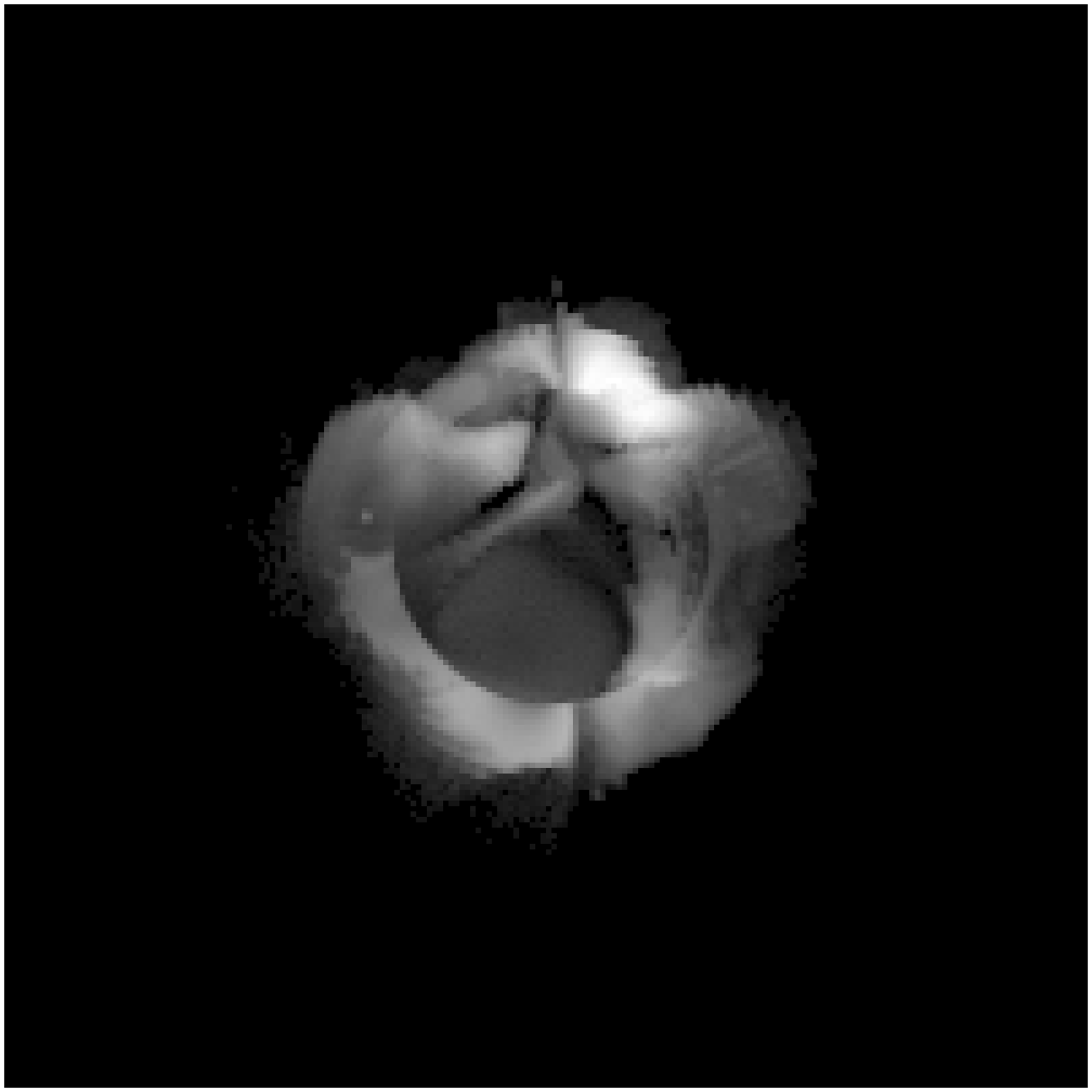,width=5.5cm}
			}  &
		\subfigure[]{
			\label{medpmod} 						
			\psfig{figure=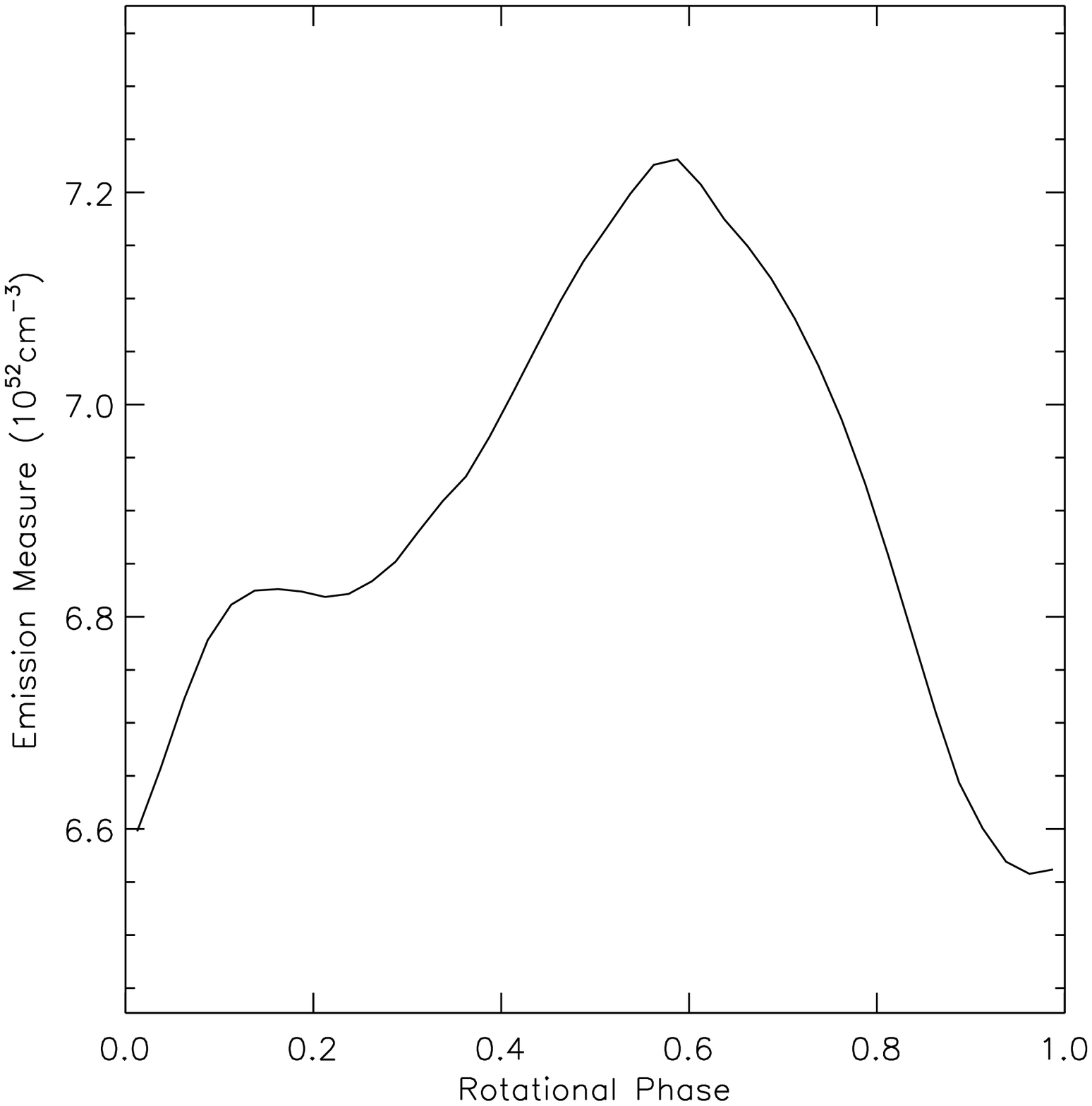,width=5.5cm}
			} \\
	\end{tabular} 
\caption[]{The effect of the coronal density on the emission measure. 
Shown are both emission measure images (viewed from longitudes
$128^{\circ}$ and $38^{\circ}$) and the corresponding rotational
modulation.  All models have the same magnetic structure as Fig. 
\ref{128_linesonly}, a temperature of 10$^{7}$K and a cutoff imposed in
the emission for $\beta>1$.  The top panels are for a model with an
emission-measure weighted density $n_{e}=4\times 10^{8}$cm$^{-3}$,
while the bottom panels have $n_{e}=1.5\times 10^{10}$cm$^{-3}$.}
    \label{images}
\end{figure*}	

\scite{jardine01structure} also examined the change in field structure
produced by a ``pseudo-dipole'' field hidden in the dark polar cap. 
This field is not a true dipole since the longest field lines (those
that emerge from the highest latitudes on the star) extend to the
source surface and so are forced to be open.  As the flux in the
dipolar component is increased, it has a progressively greater
influence on the overall field structure until, once it dominates
completely, the large-scale field is indistinguishable from the
pseudo-dipole.  Beyond this point, increasing the polar field strength
has little effect on the field structure, but it would increase the
emission measure.  Since for an isothermal corona, the density scales
as the plasma pressure (and hence in our model as the field strength
squared) we would expect the emission measure to scale as the polar
field strength to the power four.  Fig.  \ref{em_dip} shows the
variation of the emission measure with the strength of a polar field
added in to the combined surface map shown in Fig. 
\ref{mixed180_icont0}.  Adding a weak polar field ($\leq 200$G) 
reduces the emission measure as it forces some of the high-latitude 
field that was originally closed to become open. Beyond this field 
strength, the addition of the dipolar component increases 
the emission measure. Even with a polar field strength of a few
kilogauss, however, the emission measure is increased by only a factor of 10. 
Dipolar fields stronger than this would be observable in the Zeeman
maps.

Clearly, while the effect of the hidden flux is important in
influencing the field topology and the location of the open field
regions from which the wind may escape, the observational capabilities
we have at present are not sufficient to distinguish the nature of
this flux.  The different cases we have examined all show rotational
modulations less than around $20\%$ which is close to the observed
range.  While adding a dipole field could in principal give an
observable change to the emission measure, the field strengths that
can be hidden in the dark polar cap are not large enough to make an
observable difference.

In the light of this, we keep the same field topology for the rest of
this paper, using only the combined surface map shown in Fig. 
\ref{mixed180_icont0}.

\subsection{The effect of the coronal temperature}

The coronal temperature influences the X-ray emission primarily
through its effect on the density scale height.  As
Fig.~\ref{images_lowT} shows, at T=10$^{6}$K the corona is much more
confined than at T=10$^{7}$K. Lowering the temperature not only lowers
the overall emission level (in this case by two orders of magnitude)
but it increases the rotational modulation.  This increase (from $5\%$
to $12\%$) is, however, small because the regions that emit most
strongly are at high enough latitudes that they are always in view. 
In both these cases, the magnetic structure is the same.  Hence, for
this low-temperature model, large loops still exist, but their summit
density is much lower and hence they are not so bright in X-rays. 
This may of course have implications for the formation of the very
massive prominences on AB Dor that are observed at some 3-5R$^{\star}$
from the rotation axis.  If the density at this height is too low, it
may simply not be possible to sustain the radiative instability
required to form the prominences.

\subsection{The effect of the coronal density}

This is perhaps the most critical parameter to examine as it is
currently the subject of such debate.  We have chosen a sample of base
densities and calculated the total emission measure and the rotational
modulation.  In addition, we have explored the effect of modelling the
breaking open of field lines when the gas pressure exceeds the
magnetic pressure.  In these cases, we imposed a cutoff in the
emission (i.e. the density was set to zero) on those field lines where
at any point along their length the plasma $\beta$, which is the ratio
of plasma to magnetic pressure, was greater than one.  This cutoff has
little effect in models where the coronal temperature is low
(T=10$^{6}$K) since in this case the plasma pressure is also low.  For
the higher temperature models, however (T=10$^{7}$K) it makes a more
significant difference.

Fig.~\ref{images} shows emission measure images of the X-ray corona at
two different densities.  In both cases, there is a cutoff in the
emission for those fieldlines with $\beta > 1$.  Also shown is the
rotational modulation of the emission measure.  These images show
clearly that as the coronal density rises, the accompanying rise in
plasma pressure forces open more field lines and so the volume of the
{\em emitting} corona shrinks, even although the total emission
measure is rising.  This is accompanied by a slight increase in the
magnitude of the rotational modulation, although it is still well
within the observed range.  The effect of imposing a cutoff in the
emission where the plasma pressure exceeds the magnetic pressure can
be seen by comparing Figs.  \ref{nocutoff_image_1} and
\ref{lowp_image_1}.  For both of these models, the viewing angle, base
density and coronal temperature are the same, but in Fig. 
\ref{nocutoff_image_1} there is no cutoff imposed.  As a result, the
emitting volume is independent of the density and pressure and
the emission measure is larger.  The rotational modulation is small in
both cases.

In order to quantify this behaviour we have run a series of models of
increasing base density and calculated the rotational modulation and
emission measure.  Fig.  \ref{rotmod_ne} shows the rotational
modulation for a family of models at $T=10^{7}$K where a cutoff in the
emission is imposed for $\beta > 1$.  The modulation is remarkably
constant for a very wide range of coronal densities.  It only begins
to increase significantly at densities in excess of
$10^{12}$cm$^{-3}$.  At these densities, the plasma pressure at the
loop footpoints is so large that even very low-lying loops are forced
open by the plasma pressure.  As a result, the rotational modulation
increases steeply and the emission-measure weighted density in fact
falls slightly. In the case where no cutoff was imposed (or the 
temperature was so low at $T=10^{6}$K that the cutoff was never 
needed) the rotational modulation is constant at $5\% $ for all models.

Although the rotational modulation of the emission measure varies
little with the density of the emitting corona, the magnitude of the
emission measure increases steeply.  Fig.  \ref{em} shows the results
of models at $T=10^{7}$K (solid) and $T=10^{6}$K (dashed).  At each
temperature, the curves branch into two: the upper branch is the case
where there is no cutoff, while the lower branch shows how the
emission is reduced when a cutoff is imposed.  In the case where there
is no cutoff in the emission, the volume of the emitting corona is
independent of the density and so the emission measure essentially
scales as $n_{e}^{2}$.  This can be seen in the slope of the upper
branches.  As noted in the discussion of the rotational modulation, an
attempt to increase the base density much above $10^{12}$cm$^{-3}$
actually results in a reduction of the emission-measure weighted
density, as can be seen in the last few data points.

\begin{figure}
	
	\psfig{figure=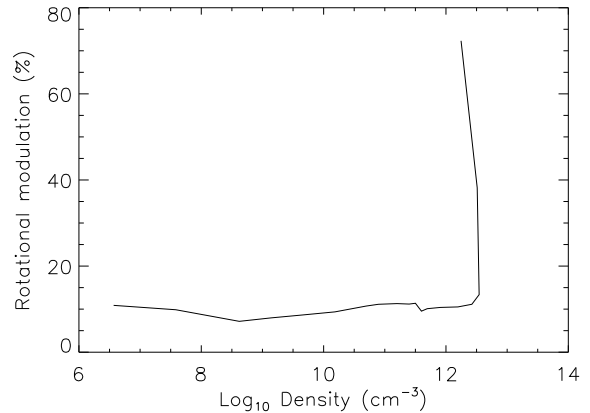,width=8.6cm}
 
	\caption[]{Rotational modulation of the emission measure as a
	function of the emission-measure weighted coronal density.}
 
	\label{rotmod_ne}
	
\end{figure}
\begin{figure}
	
	\psfig{figure=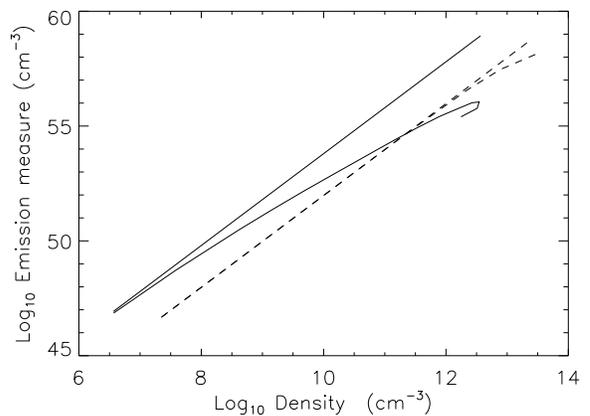,width=8.6cm}
 
	\caption[]{Total emission measure as a function of the
	emission-measure weighted coronal density.  Models with
	isothermal coronae at 10$^{6}$K (dashed) and 10$^{7}$K (solid)
	are shown.  In each case the curves have two branches. The lower 
	branch shows the amount by which the emission is reduced if
	field lines where the plasma pressure exceeds the magnetic
	pressure are assumed to have been forced open.  This simulates
	the opening up of field lines to form a stellar wind.  }
	\label{em}
	
\end{figure}
\begin{figure}
	
	\psfig{figure=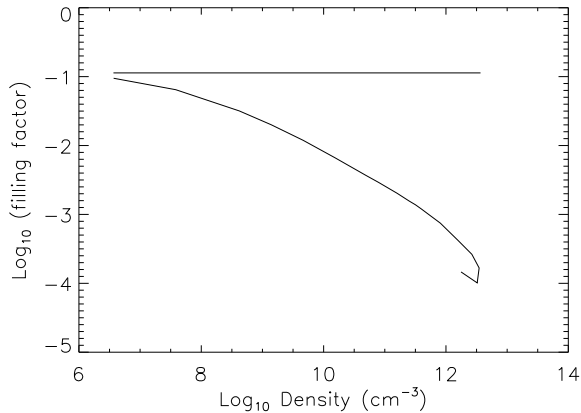,width=8.6cm}
 
	\caption[]{Density-weighted filling factor as a function of
	the emission-measure weighted coronal density.  The corona is
	assumed to be isothermal at a temperature of 10$^{7}$K. The
	lower branch shows the reduction in the filling factor if
	field lines where the plasma pressure exceeds the magnetic
	pressure are assumed to have been forced open.  }
 
	\label{fillfact}
	
\end{figure}

Clearly, cutting off the emission when the plasma pressure exceeds the
magnetic pressure results in the emission measure increasing less
steeply with density.  As the density increases, more and more field
lines reach the critical plasma pressure and the emitting volume of
the corona shrinks.  This effect can be seen in Fig.  \ref{fillfact}
where we show the behaviour of the density-weighted filling factor $f$
which we define as
\begin{equation}    
    f= \frac{\int n_{e}^{2}dV}{\frac{4}{3}(R_s^3 - R_\star^3) n_{e}^{2}}.
\end{equation} 

This is the calculated emission measure as a fraction of the emission
measure of a sphere extending to the source surface $R_{s}$ and uniformly
filled with plasma at the coronal density.  For our isothermal model,
this filling factor simply measures the fraction of the coronal volume
that is emitting.  For the case where there is no cutoff in the
emission, the filling factor is independent of the density and depends 
only on the structure of the magnetic field and particularly on the 
location and extent of the open field regions.  For the
models with a cutoff imposed, however, the filling factor falls
sharply with increasing density as a smaller and smaller fraction of
the corona contributes to the emission.  For models calculated at a
lower temperature of $10^{6}$K, the pressure scale height is smaller
and this filling factor is correspondingly less ($~2\times 10^{-4}$)
and independent of peak density.  While the behaviour of this filling
factor gives a useful qualitative understanding of the structure of
the corona of AB Dor, the numerical values should be treated with 
caution since they refer only to an isothermal
plasma.  A true stellar corona is a multi-temperature plasma, and
defining a filling factor is a much more complex issue since
fine-scale structure in both temperature and density may contribute
\cite{almleaky89,judge2000}.  Nonetheless, \scite{klimchuk2001} show 
that the commonly-used spectroscopic filling factor (similar to the 
one defined here) is a good measure of the fraction of the volume 
occupied by plasma within a narrow temperature band.

The observed range of total emission measures for AB Dor extends from 
$10^{52.8} - 10^{53.3}$cm$^{-3}$ \cite{vilhu2001}. On the basis of 
the models shown in Fig.~\ref{em} this corresponds to densities in the 
range $10^{9}-10^{10.7}$cm$^{-3}$. Using a model with a cutoff imposed 
for high-pressure fieldlines reduces this range to 
$10^{9.4}-10^{10.5}$cm$^{-3}$. In either case, the 
implied densities are high by solar standards, but span the value of 
$10^{10.5}$cm$^{-3}$ obtained from XMM-Newton observations \cite{gudel01XMM}.
In fact a model run for a temperature of $3\times 10^{6}$K 
appropriate for these observations gives an emission measure of 
$10^{53.6}$cm$^{-3}$ for a density of $10^{10.5}$cm$^{-3}$.

\section{Conclusions}

 While our calculation of the coronal magnetic field structure of AB
 Dor is based directly on the Zeeman-Doppler images, our calculation
 of the pressure and density structure of the corona is more
 model-dependent.  With the very simplest of assumptions however (an
 isothermal plasma in hydrostatic equilibrium) we have been able to
 reproduce the observed emission measures, densities and rotational
 modulation in X-rays.  The key to our ability to reconcile the high
 emission measures and densities with the low rotational modulation
 lies in the complex nature of the magnetic field.  The
 Zeeman-Doppler images show a surface that is densely covered in flux. 
 Extrapolating this surface field shows that the corona contains loops
 on all scales and with a range of field strengths.  
 In the low-density models, the corona is very extended and so shows
 little rotational modulation.  In the higher density models, the
 emitting corona is more compact, but again shows little rotational 
 modulation since the brightest regions are at
 high latitudes where they are always in view as the star rotates. 

\section{Acknowledgements}
We would like to thank Dr A. van Ballegooijen for allowing
us to use his code for calculating the potential field extrapolation.
The synoptic magnetic data used in this study were produced 
cooperatively by NSF/NOAO, NASA/GSFC, NOAA/SEL and NSO/Kitt Peak and 
made publicly accessible via the World Wide Web.




\end{document}